\documentclass{jps-cp}

\title{Field-angular Dependence of Pairing Interaction in URhGe: Comparison with UCoGe}

\author{Yo \textsc{Tokunaga}$^{1}$, Dai \textsc{Aoki}$^{2,3}$, Hadrien \textsc{Mayaffre}$^{4}$, Steffen \textsc{Kr\"{a}mer}$^{4}$, Marc-Henri\,\textsc{Julien}$^{4}$, Claude \textsc{Berthier}$^{4}$, Mladen \textsc{Horvati{\'c}}$^{4}$, Hironori \textsc{Sakai}$^{1}$, Shinsaku\,\textsc{Kambe}$^{1}$, Taisuke\,\textsc{Hattori}$^{1}$,  and Shingo \textsc{Araki}$^{2,5}$ }

\inst{$^{1}$ASRC, Japan Atomic Energy Agency, Tokai, Ibaraki 319-1195, Japan \\
$^{2}$INAC/SPSMS, CEA-Grenoble/UGA,  38054 Grenoble, France\\
$^{3}$IMR, Tohoku University, Ibaraki 311-1313, Japan\\
$^{4}$LNCMI-CNRS (UPR 3228), EMFL, UGA, UPS, INSA, 38042 Grenoble, France\\
$^{5}$Department of Physics, Okayama University, Okayama 700-8530, Japan\\}

\abst{The field-angular dependence of Co-NMR spin-lattice relaxation rate $1/T_1$ has been measured for a 10\% Co-doped single crystal of URhGe.
The experiment revealed that spin fluctuations in ferromagnetic (FM) state of URhGe are robust against magnetic field below about 4 T, applied along any direction in the $bc$ crystal plane. 
This is in clear contrast with the sister compound UCoGe, in which FM spin fluctuations are rapidly suppressed by a tiny applied field along the $c$ axis. 
We show that such a difference in the character of the spin fluctuations is reflected in their two distinct phase diagrams for the upper critical field $H_{\rm c2}$, providing further support to the mechanism of superconductivity mediated by spin fluctuations in these materials.}

\kword{Ferromagnetic superconductivity, URhGe, UCoGe}

\begin{document}
\maketitle

\section{Introduction}

In conventional superconductors,  the attractive interaction responsible for superconducting (SC) pairs is mediated by lattice vibrations (phonons). The strength of this interaction is generally assumed to be field-independent, since the lattice vibrations are not directly coupled to the applied magnetic field ($H_{\rm ext}$). The influence of $H_{\rm ext}$ is usually treated as a perturbation that results in suppression or  modulation of the SC order parameter. 
In the case of uranium-based ferromagnetic (FM) superconductors UGe$_2$, URhGe and UCoGe,  on the other hand, the pairing interaction is supposed to be mediated by FM spin fluctuations \cite{UGe2SC,AokiNature,UCoGe,Fay, Valls}, whose excitation spectrum can be strongly modified by $H_{\rm ext}$, as proved by recent NMR studies  \cite{THattori,THattori2,YTPRL,YTPRB,Kotegawa}. 
It is thus supposed that the pairing interaction can be affected by $H_{\rm ext}$, and indeed, such a field-dependent pairing mechanism has been suggested to be responsible for the unconventional SC properties of UCoGe \cite{THattori,THattori2,Mineev2011,Tada,Tada2,Wu}, 
and further, provide exotic field-induced SC in the case of URhGe \cite{YTPRL,YTPRB,Kotegawa,LevyScience,LevyNP,Hattori,Mineev2015}.

\begin{figure}[tb]
\begin{center}
\includegraphics[width=15.6 cm,keepaspectratio]{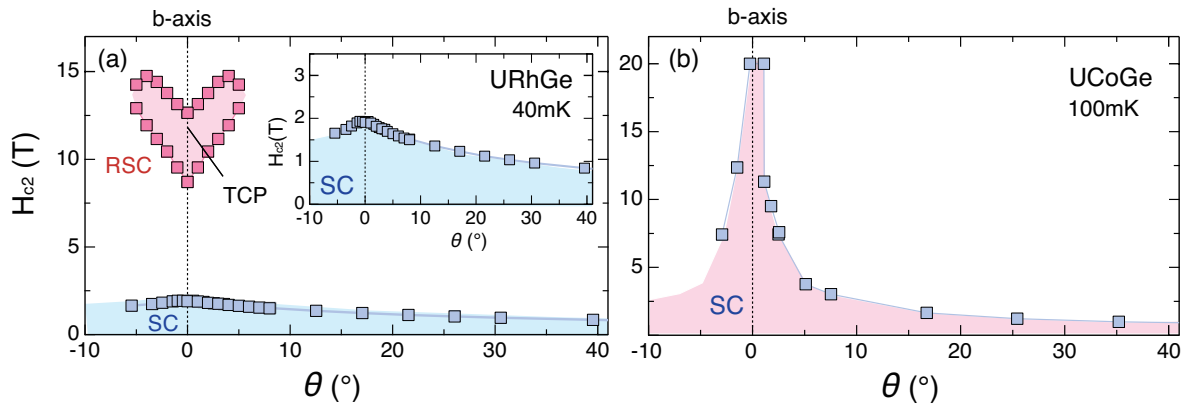}  
\caption{\footnotesize \setlength{\baselineskip}{3.5mm} Angle $\theta$ dependence of the upper critical field $H_{\rm c2}$ determined by resistivity measurements  
in single crystals of (a) URhGe and (b) UCoGe \cite{AokiReview}.}
\end{center}
\vspace{-3mm}
\end{figure}
 
Among FM superconductors, URhGe and UCoGe have
the same crystal structure of the orthorhombic TiNiSi type and are both itinerant weak ferromagnets with Ising anisotropy along the $c$ crystal axis. These two compounds should thus offer similar bases for the SC mechanism. However, their upper critical field $H_{\rm c2}$, one of the
most fundamental SC parameters, exhibits remarkable
differences\cite{Wu,Aoki2009S,AokiReview}.
In Fig.\,1, we compare the $\theta$ dependence of $H_{c2}$ between UCoGe and URhGe \cite{AokiReview}.
For UCoGe  [Fig.\,1(b)], $H_{c2}$ reaches a surprisingly large value, about 20 T for $H_{\rm ext}\|b$ ( $\sim30$ T for $H_{\rm ext}\|a$).
The value highly exceeds the Pauli limiting field, 1.84 $k_{\rm B}T_{\rm SC}/\mu_{\rm B}$, where $T_{\rm SC}=0.7$ K. 
$H_{c2}$ is, however, extremely sensitive to the angle of the applied field; a small rotation of the field from the $b$ axis to the $c$ axis rapidly suppresses the $H_{c2}$.

In contrast,  $H_{c2}$ in URhGe has no steep increase either around $H_{\rm ext}\|b$ or around $H_{\rm ext}\|a$. $H_{c2}$ exhibits only a weak $\theta$ dependence below 2 T [the inset of Fig.\,1(a)]. The anisotropy of $H_{\rm c2}$ is  only about 2.5 between the $b$ and $c$-axis directions. Such moderate anisotropy would be mostly explained by the anisotropy of the effective electron mass.
Interestingly, the field-induced ``reentrant'' SC phase (RSC) appears around a field-induced tricritical point (TCP) located around  $H_R\approx12$T for $H_{\rm ext}\|b$. At the TCP, the FM order is forced to align along the field direction ($\|b$) \cite{LevyScience,LevyNP}.

In this paper, we report the results of the Co-NMR spin-lattice relaxation rate $1/T_1$ measurements performed at lower fields for a 10\% Co-doped single crystal of URhGe.
The experiments reveal that spin fluctuations in URhGe are robust against magnetic field applied along any direction in the $bc$ plane. 
This makes a clear contrast with the case of UCoGe\cite{THattori}, and provides further support for the mechanism of FM superconductivity mediated by spin fluctuations in these materials.

\section{Experimental Results and Discussion}

A 10\% Co-doped single crystal of URhGe (=URh$_{0.9}$Co$_{0.1}$Ge) was prepared by the Czochralski pulling method.
The substitution of Co for Rh is isostructural and only  negligibly affects the magnetic properties of URhGe
 \cite{Sakarya,Huy,YTPRL}.
The compound is thus regarded as a suitable proxy to investigate the nature of FM fluctuations in URhGe \cite{YTPRL,YTPRB}. 
The $1/T_1$ measurements were performed at a fixed temperature of $T=1.6$ K, well below $T_{\rm Curie}$ = 11.8 K.
The $T_1$ values were determined by fitting the saturation recovery of the spin-echo intensity to the theoretical function for the central transition of the $I=7/2$ nuclear spins.

\begin{figure}[t]
\begin{center}
\includegraphics[width=15.8 cm,keepaspectratio]{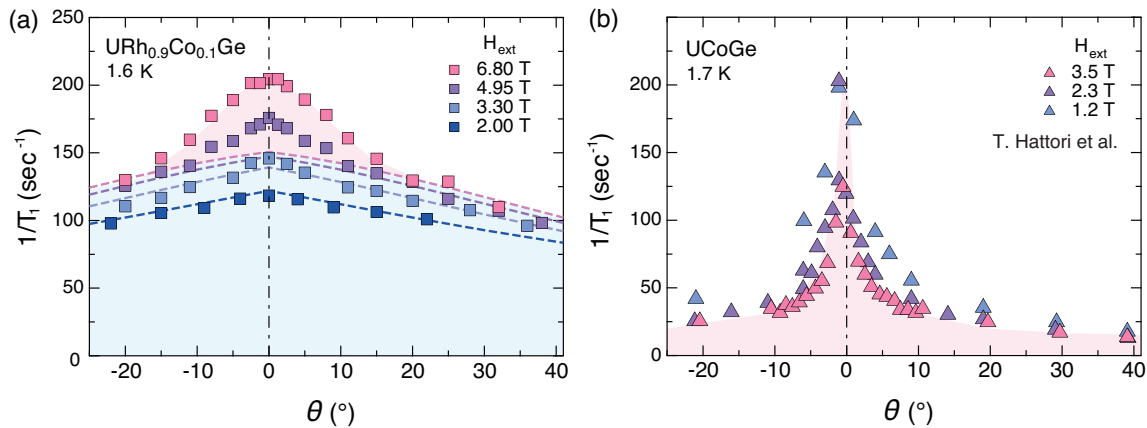} 
\caption{\footnotesize \setlength{\baselineskip}{3.5mm}  (a) The field-angular dependence of $1/T_1$ for URh$_{0.9}$Co$_{0.1}$Ge measured with several different fields, $H_{\rm ext}$=2, 3.3, 4.95 and 6.8 T. The dashed lines of the corresponding color indicate the $1/T_1(\theta)$ values calculated using the Eqs. (2) and (3) for fixed values of $(1/T_1)_b=155$ s$^{-1}$ and $(1/T_1)_c=63$ s$^{-1}$ and $H_{\rm int}=1.5$ T (see the text). 
(b) The field-angular dependence of $1/T_1$ reported for UCoGe \cite{THattori}. 
  }
\end{center}
\vspace{-3mm}
\end{figure}

Figure 2(a) shows the field-angle dependence of $1/T_1$ in URh$_{0.9}$Co$_{0.1}$Ge.
Here the angle $\theta$ is defined as  the angle from the $b$ axis in the $bc$ crystal plane; the magnetic field components are thus ($H_{\rm ext}^b$, $H_{\rm ext}^c$) = ($H_{\rm ext}\cos\theta, H_{\rm ext}\sin\theta$). $1/T_1(\theta)$ was measured for different field values, $H_{\rm ext}$=2, 3.3, 4.95 and 6.8 T. 
In Fig.\,2(b), we plot $1/T_1(\theta)$ reported by Hattori {\it et al.} for UCoGe \cite{THattori}.
In UCoGe, $1/T_1(\theta)$ exhibits a strong field-angle dependence; a small rotation of $H_{\rm ext}$ from $\theta=0^{\circ}$ dramatically suppress the $1/T_1$. This leads to a cusp-like sharp peak around $\theta=0^{\circ}$, which becomes sharper with increasing $H_{\rm ext}$.
This behavior indicates a strong suppression of the longitudinal component of FM spin fluctuations by $H_{\rm ext}^c=H_{\rm ext}\sin\theta$ \cite{THattori,THattori2,Tada,Tada2}.
It is thus naturally expected that the suppression of the pairing strength by $H_{\rm ext}^c$ is at the origin of the anomalous $H_{\rm c2}$ behavior in UCoGe \cite{THattori,Tada}.

In contrast,  we found that $1/T_1$ in URh$_{0.9}$Co$_{0.1}$Ge shows relatively weak $\theta$ and $H_{\rm ext}$ dependences at lower fields (at 2 and 3.3 T) [Fig.\,2(a)];
there is no dramatic suppression of the FM spin fluctuations by $H_{\rm ext}^c$. 
As $H_{\rm ext}$ is increased,  $1/T_1(\theta)$ variation increases, developing a broad peak centered at $\theta=0^{\circ}$. 
However, compared with the case of UCoGe, the overall $\theta$ dependence is still not very strong even above 4 T. 
In the following,  we analyze the $1/T_1$ behavior shown in Fig.\,2(a), to demonstrate that, in contrast to the case of UCoGe, the field effects on the spin fluctuations are negligibly small in URhGe at lower field values.

In general, $1/T_1$ probes hyperfine field fluctuations perpendicular to the quantization axis of the nuclear spin. 
In the paramagnetic state, the nuclear spin is quantized along the applied field direction, and hence
the $\theta$ dependence of the $1/T_1$ is expressed using the values $(1/T_1)_{b,c}$ that correspond to the fields applied along the $b$ and $c$ directions, respectively:
\begin{equation}
1/T_1(\theta)=(1/T_1)_b\cos^2\theta+(1/T_1)_c\sin^2\theta.
\end{equation}
\begin{figure}[t]
\begin{center}
\includegraphics[width=9.3 cm,keepaspectratio]{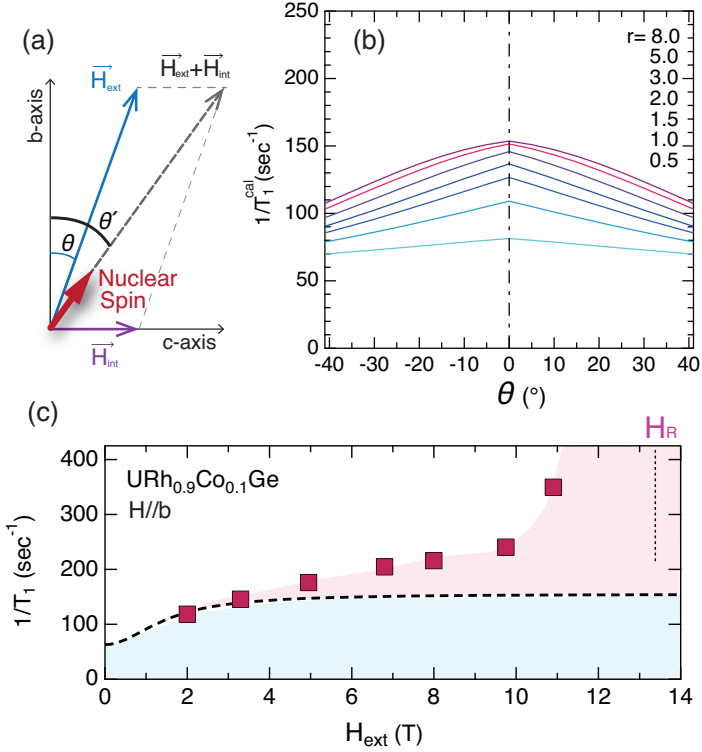} 
\caption{\footnotesize \setlength{\baselineskip}{3.5mm} (a) A schematic view of the nuclear spin quantized along the vector sum of the external and internal fields, \mbox{\boldmath $H_{\rm ext}+H_{\rm int}$}. (b) The $r$ dependence of the $1/T_1(\theta)$ calculated using the Eqs. (2) and (3) for fixed values $(1/T_1)_b=155$ s$^{-1}$ and $(1/T_1)_c=63$ s$^{-1}$. 
(c) The measured $H_{\rm ext}$ dependence of $1/T_1$ at $\theta=0^{\circ}$ ($H//b$). The dashed line is the calculated $H_{\rm ext}$ dependence using the above given $(1/T_1)_{b,c}$   and  $H_{\rm int}=1.5$ T. The $H_{\rm R}$ represents the critical field where the TCP appears. Both $1/T_1$ and $1/T_2$ diverge towards $H_{\rm R}$ \cite{YTPRL,YTPRB,Kotegawa} }
\end{center}
\vspace{-3mm}
\end{figure}
This equation was found to fully explain the $\theta$ dependence of $1/T_1$ in the paramagnetic state of UCoGe \cite{THattori}.

In the FM state, on the other hand, a typically large internal field $H_{\rm int}$ generally appears at nuclear positions.
Then, instead of being quantized along the direction of  \mbox{\boldmath $H_{\rm ext}$}, 
the nuclear spin is quantized along the vector sum of the two fields, \mbox{\boldmath $H_{\rm ext}+H_{\rm int}$} [Fig.\,3(a)]. This requires the modification of Eq. (1) as 
\begin{equation}
1/T_1(\theta)=(1/T_1)_b\cos^2\theta{'}+(1/T_1)_c\sin^2\theta{'},
\end{equation}
with
\begin{equation}
\tan{\theta{'}}=\frac{H_{\rm ext}\sin{\theta}+H_{\rm int}}{H_{\rm ext}\cos{\theta}}=\frac{\sin{\theta}+1/r}{\cos{\theta}}.
\end{equation}
These equations indicate that both the magnitude and $\theta$ dependence of $1/T_1$ depend on the ratio $r=H_{\rm ext}/H_{\rm int}$. 

In the case of  UCoGe,  $H_{\rm int}$\ has been estimated from Co-NQR \cite{Ohta} to be $\sim$0.1 T directed parallel to the $c$ axis. 
The value is more than an order of magnitude smaller than the $H_{\rm ext}$ used in NMR experiments (typically more than 1 T). This allowed the authors in the Ref. \cite{THattori} to use Eq. (1) even in the FM state.
In the case of  URh$_{0.9}$Co$_{0.1}$Ge, on the other hand, the $H_{\rm int}$ has not been determined by Co-NQR. However, the ordered moment of $0.47 \mu_B$, estimated in the FM state, is nearly 10-15 times larger than that of UCoGe, implying the existence of a large $H_{\rm int}\sim1.5$ T at Co nuclear positions.
The value is of the same order as $H_{\rm ext}$, and thus must be fully taken into account in the $1/T_1$ analysis. 

Figure\,3(b) shows the $r$ dependence of $1/T_1(\theta)$ calculated using the Eqs. (2) and (3) with fixed values of $1/T_1$, $(1/T_1)_b=155$ s$^{-1}$ and $(1/T_1)_c=63$ s$^{-1}$. 
The same calculations are also plotted in Fig.\,1(a), to be compared with the experimental data. Here, we take the $H_{\rm int}=1.5$ T, that is, for example, $H_{\rm ext}=2$ T corresponds to $r=1.3$. 
As seen in Fig.\,2(a), the two given values of $(1/T_1)_b$ and $(1/T_1)_c$ reproduce well the magnitude and $\theta$ dependence of $1/T_1$ obtained at the two low field values $H_{\rm ext}=2$ T $(r=1.3)$  and 3.3 T $(r=2.2)$.
That is,  $1/T_1$ behavior below $\sim$4 T can be explained mostly by taking into account the tilting of the quantization axis of the nuclear spins with increasing $H_{\rm ext}$.
Conversely,  it reveals that the field effect on spin fluctuations is negligibly small at the lowest fields. 

The calculation, however, does not fully reproduce $1/T_1(\theta)$ at higher fields. To be more precise, it does not explain the development of the broad peak around $\theta=0^{\circ}$ in $H_{\rm ext}=4.95$ and 6.80 T  [Fig. 2(a)].
The deviation between experiment and calculation 
becomes significant with increasing $H_{\rm ext}$. This can be seen more clearly in Fig.\,3(c), where the calculated field dependence of $1/T_1$ at $\theta=0^{\circ}$ ($\|b$) is compared with the experimental data.
While the calculated value is nearly field independent, the experimental data rapidly increase above $H_{\rm ext}\sim$8 T \cite{YTPRL}.  
As discussed in previous papers \cite{YTPRL,YTPRB}, this rapid increase of $1/T_1$ is connected to the enhancement of spin fluctuations when approaching the TCP around 13 T. 
Around the TCP,  the spin fluctuations diverge both along the $b$ and the $c$ axes, providing the divergence in $1/T_2$ and $1/T_1$, respectively \cite{YTPRL,YTPRB,Kotegawa,Misawa1,Misawa2}. 

It should be remarked that the critical fluctuations in UCoGe are suggested to develop as a feature of the system close to the FM instability \cite{THattori,AokiJPSJ83,AokiReview}. 
As mentioned, these critical fluctuations are in longitudinal mode ($H\|c$) and very sensitive to the $H_{\rm ext}$ applied along the same direction (i.e., parallel to the easy-magnetization axis) \cite{THattori,Tada}.  
On the other hand, the  present NMR results reveal that URhGe at zero field does not possess such critical fluctuations; in this system the FM instability  
and related critical fluctuations emerge only under strong magnetic fields, when the system approaches to the TCP \cite{AokiJPSJ83,AokiReview,Nakamura2017,YTPRL,YTPRB}.
Then, the higher $T_{\rm SC}$ in UCoGe (0.7\,K) and in the high-field reentrant SC phase of URhGe (0.45 K), as compared  to $T_{\rm SC}$ in URhGe at zero field (0.25 K), implies that these critical fluctuations are more favorable for the mechanism of the FM superconductivity.

\section{Summary}

In summary, the field-angle resolved NMR $1/T_1$ measurements reveal that spin fluctuations in the FM state of URhGe are robust against moderate magnetic field ($\lesssim4$ T) applied along any direction in the $bc$ crystal plane. 
This makes a clear contrast with the case of its sister compound UCoGe, in which the rapid suppression of the FM spin fluctuations by a tiny field along the $c$ axis has been observed. 
We show that such a difference in the character of spin fluctuations is reflected in their two distinct phase diagrams of the upper critical field $H_{\rm c2}$. This provides further support for the mechanism of superconductivity mediated by FM spin fluctuations in these materials.

\section*{Acknowledgment}
A part of this work was supported by JSPS KAKENHI Grant Number JP15KK0174, JP15H05745, JP15H05884, JP15H05882, JP15K21732, JP16KK0106, JP16H04006, 19K03726, JP19H00646, ICC-IMR, and the REIMEI Research Program of JAEA.

\end{document}